# Insight into the Distribution of High-pressure Shock Metamorphism in Rubble-pile Asteroids

Nicole Güldemeister[1,2], Juulia-Gabrielle Moreau[3], Tomas Kohout[1,4], Robert Luther[2], and Kai Wünnemann[2,5]
[1] Department of Geosciences and Geography, University of Helsinki, Finland; tomas.kohout@helsinki.fi
[2] Museum für Naturkunde, Leibniz Institute for Evolution and Biodiversity Science, Berlin, Germany; ngueldemeister@gmx.de
[3] Institute of Ecology and Earth Sciences, Department of Geology, University of Tartu, Estonia
[4] Institute of Geology of the Czech Academy of Sciences, Prague, Czech Republic
[5] Freie Universität Berlin, Institut für Geologische Wissenschaften, Germany
*Received 2022 May 5; revised 2022 July 22; accepted 2022 July 23; published 2022 August 23*

## Abstract

Shock metamorphism in ordinary chondrites allows for reconstructing impact events between asteroids in the main asteroid belt. Shock-darkening of ordinary chondrites occurs at the onset of complete shock melting of the rock (>70 GPa) or injection of sulfide and metal melt into the cracks within solid silicates (∼50 GPa). Darkening of ordinary chondrites masks diagnostic silicate features observed in the reflectance spectrum of S-complex asteroids so they appear similar to C/X-complex asteroids. In this work, we investigate the shock pressure and associated metamorphism pattern in rubble-pile asteroids at impact velocities of 4–10 km s$^{-1}$. We use the iSALE shock physics code and implement two-dimensional models with simplified properties in order to quantify the influence of the following parameters on shock-darkening efficiency: impact velocity, porosity within the asteroid, impactor size, and ejection efficiency. We observe that, in rubble-pile asteroids, the velocity and size of the impactor are the constraining parameters in recording high-grade shock metamorphism. Yet, the recorded fraction of higher shock stages remains low (<0.2). Varying the porosity of the boulders from 10% to 30% does not significantly affect the distribution of pressure and fraction of shock-darkened material. The pressure distribution in rubble-pile asteroids is very similar to that of monolithic asteroids with the same porosity. Thus, producing significant volumes of high-degree shocked ordinary chondrites requires strong collision events (impact velocities above 8 km s$^{-1}$ and/or large sizes of impactors). A large amount of asteroid material escapes during an impact event (up to 90%); however, only a small portion of the escaping material is shock-darkened (6%).

*Unified Astronomy Thesaurus concepts:* Main belt asteroids (2036); Computational methods (1965); Asteroid dynamics (2210); Chondrites (228)

## 1. Introduction

Shock metamorphism in ordinary chondrites is a consequence of collision events between their parent bodies—asteroids. The degree of shock metamorphism (Stöffler et al. 2018) is dependent on the amount of energy (i.e., pressure) deposited upon impact as well as on material physical state (e.g., porosity). The material subjected to shock experiences mechanical (fracturing, deformation of crystalline lattice) as well as thermal (heating, partial or complete melting) alteration, and relocation (burial, ejection). Upon significant disruption of large asteroids (Asphaug et al. 1998; Michel et al. 2002), the fragmentation remnants may form asteroid families within the main asteroid belt.

The most widely adopted technique to remotely characterize and classify asteroids is based on their reflectance spectra. DeMeo et al. (2009) define two main asteroid complexes. The S-complex features strong 1 and 2 $\mu$m absorption bands, diagnostic for olivine and pyroxene, while the C/X-complex typically has a featureless monotonous spectrum, sometimes with the presence of weak absorptions around 1 or 0.7 $\mu$m. The S-complex asteroid spectrum resembles that of ordinary chondrites, which are composed of silicates with a variable content of metallic iron and iron sulfides as the main mineral phases (Norton 2002). The C/X-complex asteroid spectrum, with its weak absorptions and high albedo variation, is often linked to carbonaceous chondrites (low albedo) or metal-rich material (high albedo). However, some so-called shock-darkened ordinary chondrites with associated melting and redistribution of metal and sulfides (Heymann 1967; Britt & Pieters 1989; Britt et al. 1989; Stöffler et al. 1991, 2018; Keil et al. 1992; Britt & Pieters 1994; Kohout et al. 2014, 2020) are characterized by optical change toward a dark and featureless spectrum being a remarkably good fit to spectra of the C/X-complex asteroids. Thus, if an asteroid fragment of S-complex asteroids is shock-darkened, its spectra may be indistinguishable from those of C/X-complex asteroids. Such an observation would affect the generally accepted distribution of asteroid compositions in the main asteroid belt.

As observed by Kohout et al. (2020) in a shock experiment with the Chelyabinsk meteorite (porosity 6%), shock-darkening happens at two different pressure ranges of the shock scale at shock stage C-S7 (>70 GPa, whole rock melting, as in Stöffler et al. 2018; >90/150 GPa, based on shock models in Figure 1(e) of Kohout et al. 2020) or between shock stages CS-5 and 6 (40–60 GPa, metal and iron sulfide melt veins; Kohout et al. 2014; Moreau et al. 2017, 2018, 2019a; Moreau 2019; Moreau & Schwinger 2021). The upper limit of the 40–60 GPa pressure interval is given by the onset of silicate melting and immiscibility of troilite and silicate melts, preventing infusion of troilite melt into residual grains. At higher porosity (15%–30%), the corresponding pressures for shock-darkening will be lowered by approximately 10 GPa (Moreau et al. 2019a) to 30–40 GPa and >60 GPa.

As we see from these conditions for shock-darkening, the porosity and intensity of the impact on the asteroid are major factors determining the mass of shock-darkened material. The intensity of the shock is a direct function of impact velocity, density, and porosity of the impactor, and impact obliquity. However, the final distribution of shock pressures in the impacted asteroid will strongly depend on its internal structure (Davison et al. 2010, 2012). Asteroids can either be monolithic, with homogeneous distribution of the internal porosity, or rubble-pile asteroids (Taylor et al. 1987; Britt et al. 2002), where porosity is heterogeneously distributed between boulders and finer materials. Such rubble-pile asteroids are believed to be a collection of rocks loosely bound by their own gravity or cohesive forces. Their accretion is assumed to be accompanied by particle size sorting. Boulders are surrounded by fine porous material or voids, where larger voids are also located near the center of the asteroid (Wada et al. 2018), resulting in a heterogeneous, highly porous structure. Therefore, an impact of similar intensity on a rubble pile may have a different, more heterogeneous outcome than on a monolithic asteroid.

To reveal and quantify shock metamorphism in rubble-pile asteroids and to estimate the fraction of shock-darkened material, we carry out a numerical study of impact modeling considering targets with internal structures similar to a rubble-pile asteroid, and vary the impactor velocity and size as well as their internal porosity patterns. Furthermore, we trace any ejected shock-darkened material and determine the ratio of ejected to remaining shock-darkened material for each scenario.

## 2. Methods

To simulate impact processes, we used the iSALE shock physics code (Wünnemann et al. 2006; Elbeshausen et al. 2009; Elbeshausen & Wünnemann 2011) with custom modifications from Moreau et al. 2019b) for the iSALE-2D code (use of BMP files as material map input). The iSALE-2D code is based on the SALE hydrocode solution algorithm (Amsden et al. 1980). To simulate hypervelocity impact processes in solid materials, SALE has been modified to include an elastic–plastic constitutive model, fragmentation models, various equations of state, and multiple materials (Melosh et al. 1992; Ivanov et al. 1997). More recent improvements include a modified strength model (Collins et al. 2004) and a porosity compaction model (Wünnemann et al. 2006; Collins et al. 2011). The iSALE-3D code uses a solver as described in Hirt et al. (1974). iSALE has been benchmarked against other hydrocodes (Pierazzo et al. 2008) and validated against experimental data from laboratory-scale impacts (Pierazzo et al. 2008; Davison et al. 2011; Miljkovic et al. 2012; Güldemeister et al. 2013). iSALE shock physics code can also accurately model the ejection process as indicated by a detailed comparison between models and NASA Ames experimental data for impacts on quartz (Güldemeister et al. 2015; Wünnemann et al. 2016; Luther et al. 2018; Raducan et al. 2019). The suitability of model ejecta has also been shown for large-scale cratering (Zhu et al. 2015, 2017).

### 2.1. Projectile and Target Material Properties

For all presented models, we used the analytical equations of state (Thompson & Lauson 1972) for dunite (Benz et al. 1989)

**Table 1**
Strength and Porosity Properties (Dunite)

| Strength Parameters in iSALE[a] | |
| --- | --- |
| Strength (intact) $Y_{i0}$ | 50 MPa |
| Strength (damaged) $Y_{id}$ | 0.05 MPa |
| Limited strength (intact) $Y_{ilim}$ | 3500 MPa |
| Limited strength (damaged) $Y_{idam}$ | 3000 MPa |
| Coefficient of internal friction (intact) $\mu_i$ | 1.2 |
| Coefficient of internal friction (damaged) $\mu_d$ | 0.6 |
| Porosity Parameters in iSALE | |
| Distension $\alpha = 1/(1-\Phi_{porosity})$ | 1.11/1.43/4.0 |
| Compaction rate $\kappa$ | 0.98 |
| Speed of sound ratio from porous to solid material ($\chi$) | 1 |
| Distension to power law $\alpha_X$ | 1.00 |
| Volume strain at plastic compaction, elastic threshold $\varepsilon_e$ | $-1 \ast 10^{-5}$ |

**Note.**
[a] From Cremonese et al. (2012)

as provided by the iSALE shock physics code. It helps to simulate impacts on materials of silicate composition (olivine $Fo_{90}$, pyroxene) as a proxy for ordinary chondrites (Moreau et al. 2017, 2018, 2019a; Moreau 2019). Further, we used the $\varepsilon$–$\alpha$ compaction model (Wünnemann et al. 2006; Collins et al. 2011) to account for the porosity of the projectile, boulders, and fine material. To simplify our study and reduce the number of parameters that can influence the final results, we applied singular strength properties regardless of the model setup and porosity. Using the strength and damage model from Collins et al. (2004), we applied properties (see Table 1) from Cremonese et al. (2012), who simulated impacts on asteroids. The absolute value of the elastic threshold $\varepsilon_e$ used in the porosity models is chosen to be very low, which allows for an immediate crushing of pore space.

### 2.2. Model Resolution

We applied a resolution (cells per projectile radius, CPPR) of a minimum 80 CPPR (Moreau 2019) based on the volume of target material shocked between 40 and 50 GPa (Pierazzo et al. 1997; Wünnemann et al. 2008; Moreau 2019) determined in a test run with a 0.8 km dunite projectile impacting a 2.0 km dunite target at a velocity of 4 km s$^{-1}$. The error for 80 CPPR, based on an extrapolation between 40 and 100 CPPR, is 2.7% (see Figure 11 in Moreau et al. 2019a).

### 2.3. Rubble-pile Model Geometry

We use numerical models of a simplified rubble-pile structure of an asteroid with a diameter of 5 km, where porosity is heterogeneously distributed between boulders of different sizes. First, larger boulders were equally distributed within the asteroid, and then smaller boulders were inserted to fill remaining space, resulting in model structure depicted in Figure 1. The structure is fully arbitrary and does not follow any actual rubble pile internal structure models. The projectile and target consist of dunitic material. The impact velocity ranges from 4 to 10 km s$^{-1}$, which covers the range of impact velocities in the main asteroid belt due to mutual collisions, which is currently estimated to be from 1 to 12 km s$^{-1}$ with a mean value of 5.3 km s$^{-1}$ (Bottke et al. 1994). The projectile diameter ranges from a small diameter of 800 m up to a large diameter of 1600 m. The porosity of the projectile is 10% and is constant throughout the parameter study. An additional study also considers a larger projectile porosity of 30%. The porosity of the target spherical boulders is 10%–30%, and they range from 150 to 850 m in size within the asteroid. The surrounding loose (fine matrix) material has a porosity ranging from 75% to 100%, where 100% corresponds to void. The model configurations used in this work are summarized in Table 2.

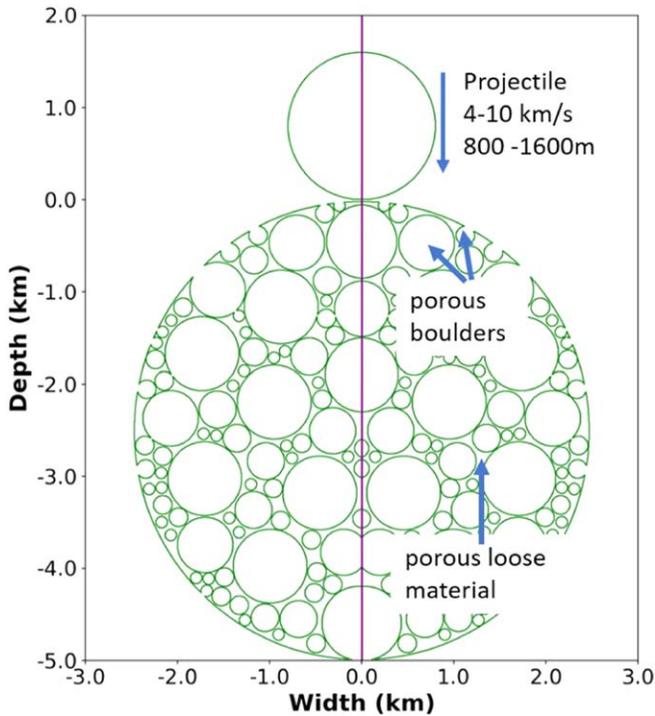

**Figure 1.** Setup of a modeled rubble-pile asteroid with variation of target and projectile properties. Porosity is heterogeneously distributed (macroporosity) and ranges from 10% to 30% for the boulders of various sizes and from 75% to 100% for the surrounding loose material.

The described models are specifically designed to study the pressure distribution within the target, how each of the parameters influences the final distribution, and how they correlate with shock stages that are required for shock-darkening. In addition to the rubble-pile asteroid scenario, we also present a reference model of a monolithic asteroid with its total porosity (in this case in the form of microporosity) equal to the averaged porosity fraction of boulders and finer material in the rubble-pile asteroid models. For instance, a boulder porosity of 10% combined with a porosity of 75% for the loose material would correspond to a total porosity of 24% considering a homogeneous distribution of porosity (monolithic asteroid), which lies within the range of asteroid porosities as described in Britt et al. (2002). S-type asteroids, hosting grain densities of 3.5–3.7 kg m$^{-3}$, are characterized by a large range of bulk densities (and thus total porosities), with an average of 2.7 kg m$^{-3}$, and porosities ranging from <10% (Vesta, Massalia) to 35% (Ida) and >50% (Britt et al. 2002; Carry 2012) with an average porosity of 25%–30%.

The reference models are meant to evaluate whether the distribution of peak shock pressures in monolithic asteroids is similar to that of rubble-pile asteroids of similar bulk porosity, as this could potentially simplify the model setup and reduce the complexity of the modeled asteroid structure.

### 2.4. Analysis of Ejected Material

Previous studies have shown that material models implemented in iSALE accurately reproduce experimental results in terms of crater size and material excavation (e.g., Güldemeister et al. 2015; Wünnemann et al. 2016; Luther et al. 2018; Winkler et al. 2018; Raducan et al. 2019). However, the analysis of ejected material requires some care because the ejecta curtain tends to be resolved by only a few cells. Here, we follow the approach of Luther et al. (2018) and determine the ejection characteristics at a specified altitude close above the surface (20 cells). When tracers that initially are equally distributed in each material-filled cell exceed this altitude, we record their velocities, ejection angles, horizontal locations, and the masses that belong to these tracers. For this study, we adjusted the ejection criterion to account for a spherical target (Figure 2). We construct a spherical envelope with a radius that is slightly bigger (plus 20 cells) than the asteroid's radius, according to the altitude criterion, to identify ejected material. When a tracer intersects the envelope, we record the ejection angle with respect to the local tangent to the spherical envelope. The ejection position is calculated by extrapolation of the velocity vector to the surface of the asteroid. To determine the amount of ejecta escaping from the system with respect to the total ejecta, we compare the local normal velocity of the ejecta with the escape velocity of the system.

When the radial velocity of the tracer exceeds the escape velocity, we assume that the tracer is escaping. The escape velocity for each tracer has been determined by using $v_{\rm esc} = \sqrt{((2 * G * M_{\rm ast})/r_{\rm centre})}$, where $G$ is the gravitational constant, $M_{\rm ast}$ the mass of the asteroid in kg, and $r_{\rm center}$ the distance of the respective tracer to the center of the asteroid in meters. Thus, the escape velocity is dependent on gravity (distance to asteroid surface). The obtained average escape velocity in our case is $\sim 0.77$ m s$^{-1}$. The mass represented by the ejected tracer equals the mass of the material within the initial cell where the tracer was originally located in. Then, we sum up all the masses of tracers which have been escaping.

## 3. Results

In this section, we first present the pressure and porosity evolution during a specific collision event (Section 3.1), followed by the effect of collision parameters on the pressure distribution (Section 3.2). The determined pressures are converted into shock stages according to the shock classification after Stöffler et al. (2018), considering a range of collision parameters as presented in Section 3.3. The efficiency of generating shock-darkening in rubble-pile asteroids for different collision scenarios and quantitative estimation of shock-darkened mass fraction is determined in Section 3.4. In Section 3.5, we compare our rubble-pile results with those of a monolithic asteroid with homogeneously distributed porosity. The results are further supplemented in Section 3.6 by the quantification of ejected shock-darkened material.

### 3.1. Pressure and Porosity Evolution

The spatial distribution of peak pressure and porosity changes as a result of the collision process is significantly dependent on the impact scenario. Figure 3 presents an overview of the evolution of peak shock pressures (left panel) and porosity (right panel) during a collision of a 1600 m large impactor of 10% porosity into a 5 km rubble-pile asteroid with boulders of 10% porosity surrounded by finer material of 75% porosity with an impact velocity of 10 km s$^{-1}$. For better visibility, Figure 3(a) represents tracers that are set back to their original position, where they recorded the respective porosities and pressures, whereas in Figure 3(b) we show the final collapsed position of the material. Generally, the collision event is a very destructive process, as seen in the last snapshot, with almost the entire impacted asteroid deformed and large amounts of material either compressed or ejected. The porosity plots clearly show that the fine or loose material is completely crushed and the boulders are highly compacted. The observed crushing behavior is independent of the porosity of the boulder or matrix. The peak shock pressures are heterogeneously distributed. In rubble-pile asteroids, the shock wave passes through different materials of various impedances (i.e., porosity), leading to a heterogeneous distribution of pressures within the boulders and the loose material (Figure 3). Reflections and superpositions at the boundaries also lead to increased pressures and a heterogeneous distribution of pressures. The largest pressures reach values above 150 GPa at the impact point, where the initial shock wave passes through the material. Pressures then decrease with increasing depth, with half of the asteroid volume still reaching pressures above 20 GPa.

Table 2
Model Configurations Used in This Work

| Target Type | Impact Velocity (km s$^{-1}$) | Impactor Diameter (m) | Target Porosity Matrix (%) | Target Porosity Boulders (%) | Target Bulk Porosity (%) |
| --- | --- | --- | --- | --- | --- |
| Rubble-pile | 4 | 800, 1200, 1600 | 75 | 10, 30 | 24, 40 |
| Rubble-pile | 4 | 800, 1200, 1600 | 100 | 10, 30 | 30, 45 |
| Rubble-pile | 6 | 800, 1200, 1600 | 75 | 10, 30 | 24, 40 |
| Rubble-pile | 6 | 800, 1200, 1600 | 100 | 10, 30 | 30, 45 |
| Rubble-pile | 8 | 800, 1200, 1600 | 75 | 10, 30 | 24, 40 |
| Rubble-pile | 8 | 800, 1200, 1600 | 100 | 10, 30 | 30, 45 |
| Rubble-pile | 10 | 800, 1200, 1600 | 75 | 10, 30 | 24, 40 |
| Rubble-pile | 10 | 800, 1200, 1600 | 100 | 10, 30 | 30, 45 |
| Monolithic | 4 | 800, 1600 | … | … | 24, 40 |
| Monolithic | 8 | 800, 1600 | … | … | 24, 40 |

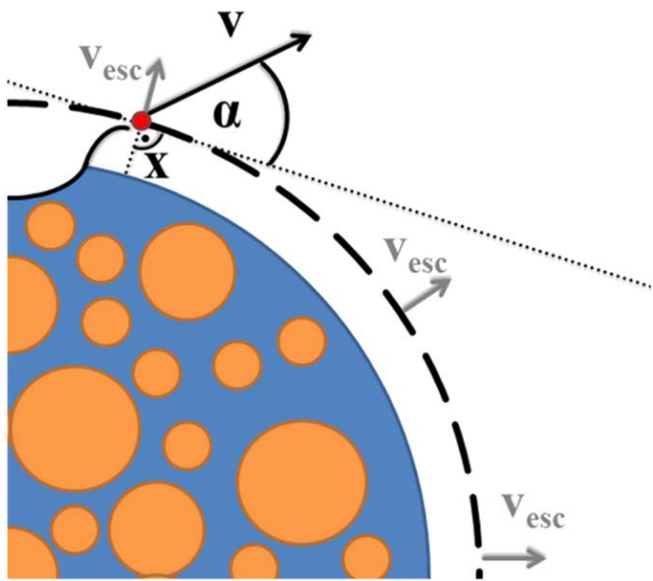

**Figure 2.** Sketch of the material ejection from a spherical heterogeneous asteroid (matrix in blue, boulders in orange). The ejection altitude criterion is represented by the black dashed line. The dotted lines respectively show the tangential and normal lines to the criterion circle for the ejected red particle at distance $x$ on the circle. The ejection angle is measured against the local tangent. The ejection velocity is indicated by a black arrow. The escape velocity is indicated by gray arrows and in a normal direction for different positions. Note that the launch distance is constructed from the extrapolation of the velocity vector to the asteroid surface.

### 3.2. Effect of Collision Parameters on Pressure Distribution

All models show a decay of pressures with depth and a heterogeneous distribution of pressures, due to the impedance contrasts between boulders and fine material affecting the outcome of a high shock metamorphism pattern within the boulders. Thus, the distribution of peak shock pressures strongly depends on both impact velocity and projectile size (Figures 4(a), (b)). Scenarios with cases involving higher boulder porosities (30%) only result in a small decrease in pressure (Figure 4(c) compared with Figure 4(a); left panels). Porosity is responsible for overall energy absorption and can lower the pressures the boulders experience, thus lowering the corresponding shock stages. It also leads to localized pressure amplifications and small volumes of material that reach higher shock stages. Asteroid macroporosity represented by empty voids (Figure 4(c), right panel) does not lead to any significant differences compared with the case of macroporosity represented by a 75% porous matrix (Figure 4(a), left panel).

### 3.3. Effect of Collision Parameters on Shock Stages

The peak shock pressures determined in Section 3.2 can be converted into shock stages after Stöffler et al. (2018). Thus, we determined the apparent fraction of each constituent in the asteroid that experienced a certain shock stage as illustrated in Figure 5 (boulders on left, loose material on right). The fractions are shown for different impact velocities ranging from 4 to 10 km s$^{-1}$. The darker the areas, the higher the shock stages. Shock stages C-S5 to C-S7 are required to generate shock-darkening in the asteroid materials by considering pressure ranges between 40 and 60 and above 70 GPa for shock-darkening. In general, the boulders experience larger fractions of specific shock stages than the loose material. In all shown collision scenarios, the fraction for certain shock stages significantly increases with increasing impact velocity. A decrease in impactor size (Figure 5) leads to a strong decrease in shock stages reached. Only for the high impact velocities do significant fractions of material experience shock stages that would lead to the formation of shock-darkening. Increasing the boulder porosity from 10% to 30% (Figure 5(c)) slightly decreases the amount of shock-darkened material, although we see that most shock-darkened material falls into the range of 40–60 GPa. Thus, the amount of material that reaches the C-S7 shock stage is not significantly affected by the change in boulder porosity. Overall, we observe a strong increase in fractions for the largest impact velocity. The change in porosity for the loose material (Figure 5(d)) from 75% to void (empty space) does not result in any significant differences in the fraction of shocked material. We kept an impactor of 1600 m diameter, to compare the two latter cases (Figures 5(b), (c)) with the first case shown in Figure 5(a).

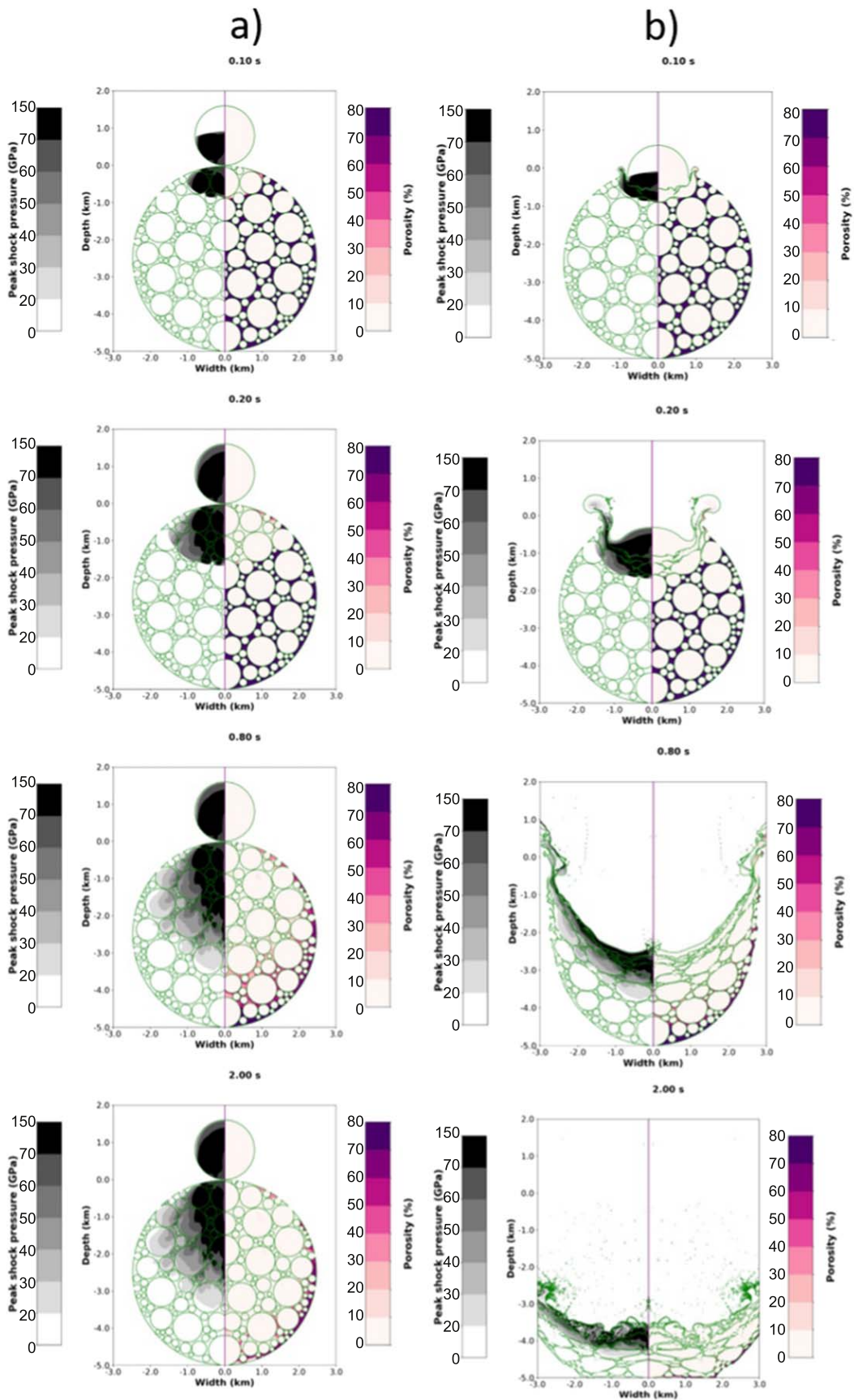

**Figure 3.** Snapshot series of the evolution of pressure (left panels) and porosity (right panels) during the collision of a 1600 m diameter projectile impacting with 10 km s$^{-1}$ velocity into a 5 km diameter rubble-pile asteroid. (a) Tracers are displayed at their initial position for better visibility of pressure and porosity values. (b) Tracers are displayed at their current position at a fixed time after impact.

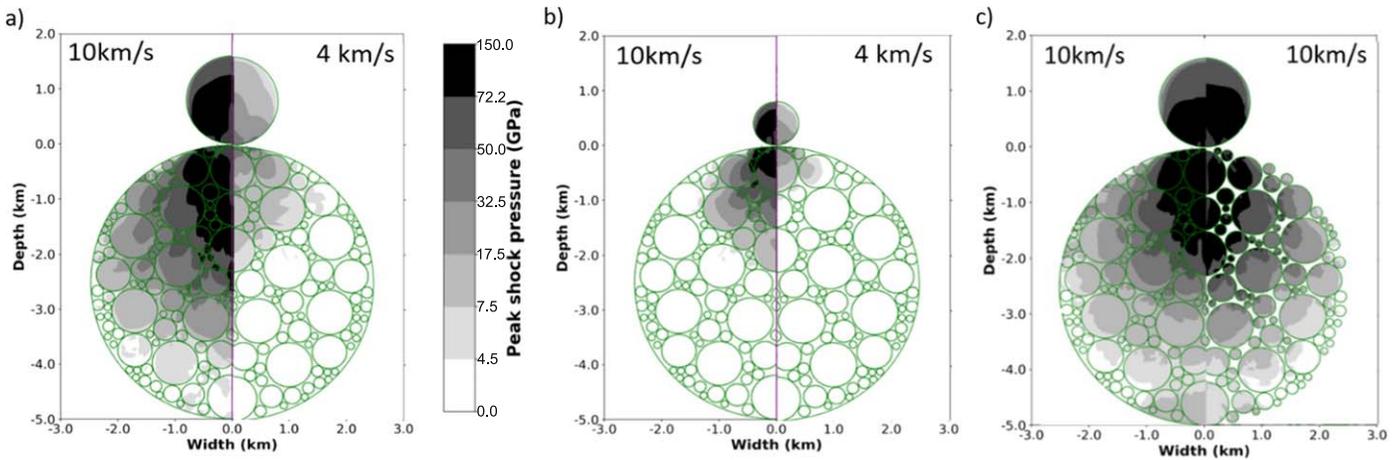

**Figure 4.** Pressure distribution after collision of an impactor of 10% porosity into a 5 km rubble-pile asteroid with boulders surrounded by fine matrix material. In (a) and (c), the projectile has a diameter of 1600 m, in (b) 800 m. The matrix porosity is 75% in (a) and (b), as well as in the left panel of (c). The boulder porosity is always 10%, except in the left panel of (c), where it is 30%. The matrix in the right panel of (c) is void (100% porosity). The respective impact velocities are noted on each subplot. Thus, (a) shows a comparison of two different impact velocities, in (b) the same scenario is presented for decreased impactor size, and in (c) the boulder porosity is increased in the left panel, whereas in the right panel, the matrix porosity is increased (void). In all figures, the tracers recording the pressure have been set to their initial position, for better visibility.

### 3.4. Efficiency of Shock-darkening

So far, we have demonstrated that shock pressures can be reached that are able to generate shock-darkening. The question remains how efficient (or abundant) shock-darkening is in rubble-pile asteroids, considering the range of modeled collision scenarios. We define the efficiency of shock-darkening as the shock-darkened mass in the rubble-pile asteroid normalized to the projectile mass ($m_t/m_p$). To quantify the efficiency, we calculate the shock-darkened mass in the asteroid by using the sum of the masses of all tracers that have experienced pressures between 40 and 60 GPa and above 150 GPa (shock-darkening regions from Kohout et al. 2020) and normalize the resulting mass to the projectile mass. From Figure 5, we can observe that the majority of the shock-darkened material is in the lower interval of 40–50 GPa. Lowering the upper interval boundary from the more conservative value of 150 GPa down to 70 GPa estimated by Stöffler et al. (2018) will result in an increase of only a few percent in shock-darkened material volume, and thus, we can keep the more conservative 150 GPa boundary in further analysis. Note that, in cylindrical symmetry for two-dimensional models, the mass of a cell or tracer is represented by the mass of a ring. As a large number of boulders and materials are considered, the assumption of taking the sum of all ring masses is a very good approximation. As shown in Figure 6, the efficiency of shock-darkening increases significantly with increasing impact velocity and is dependent on the porosity of the boulders (compare the square symbols for 10% boulder porosity with the triangular ones for 30% boulder porosity). A high boulder porosity of 30% leads to a decrease in the required shock-darkening pressures and thus an increase in shock-darkening efficiency. The mass of the asteroid is $1.6 \cdot 10^{14}$ kg, in the case of 10% boulder porosity and 75% surrounding fine material porosity, and $1.3 \cdot 10^{14}$ kg for 30% boulder porosity. Thus, for the most energetic impacts (large impactor and large impact velocities of $10\,\mathrm{km\,s^{-1}}$) an efficiency of 8 is reached, which corresponds to about 30% of the asteroid being shock-darkened. For a smaller impactor, this would only lead to 4% of asteroids being shock-darkened. In another example, considering a 1200 m impactor with an impact velocity of $10\,\mathrm{km\,s^{-1}}$, an efficiency of 5 is reached, corresponding to 8% of the asteroid being shock-darkened. Increasing the porosity of the boulders from 10% to 30% does not significantly change the efficiency. However, the effect remains stronger for a larger impact velocity. By decreasing the impact velocity, the efficiency decreases linearly. For the lowest considered impact velocities, shock-darkening is not efficient anymore.

### 3.5. Rubble-pile and Monolithic Asteroids

One aspect highlighted in the Introduction is a numerical modeling test with a simpler monolithic asteroid scenario as a proxy for collisions of rubble-pile asteroids. Therefore, in addition to the previously shown results focusing on a rubble-pile asteroid, we also consider a monolithic asteroid where porosity is homogeneously distributed and equals the average porosity of the boulders and fine material of the rubble-pile asteroid. Thus, for 10% boulder porosity and 75% surrounding matrix porosity, the resulting equal porosity of a monolithic asteroid defined within the numerical model is 24%.

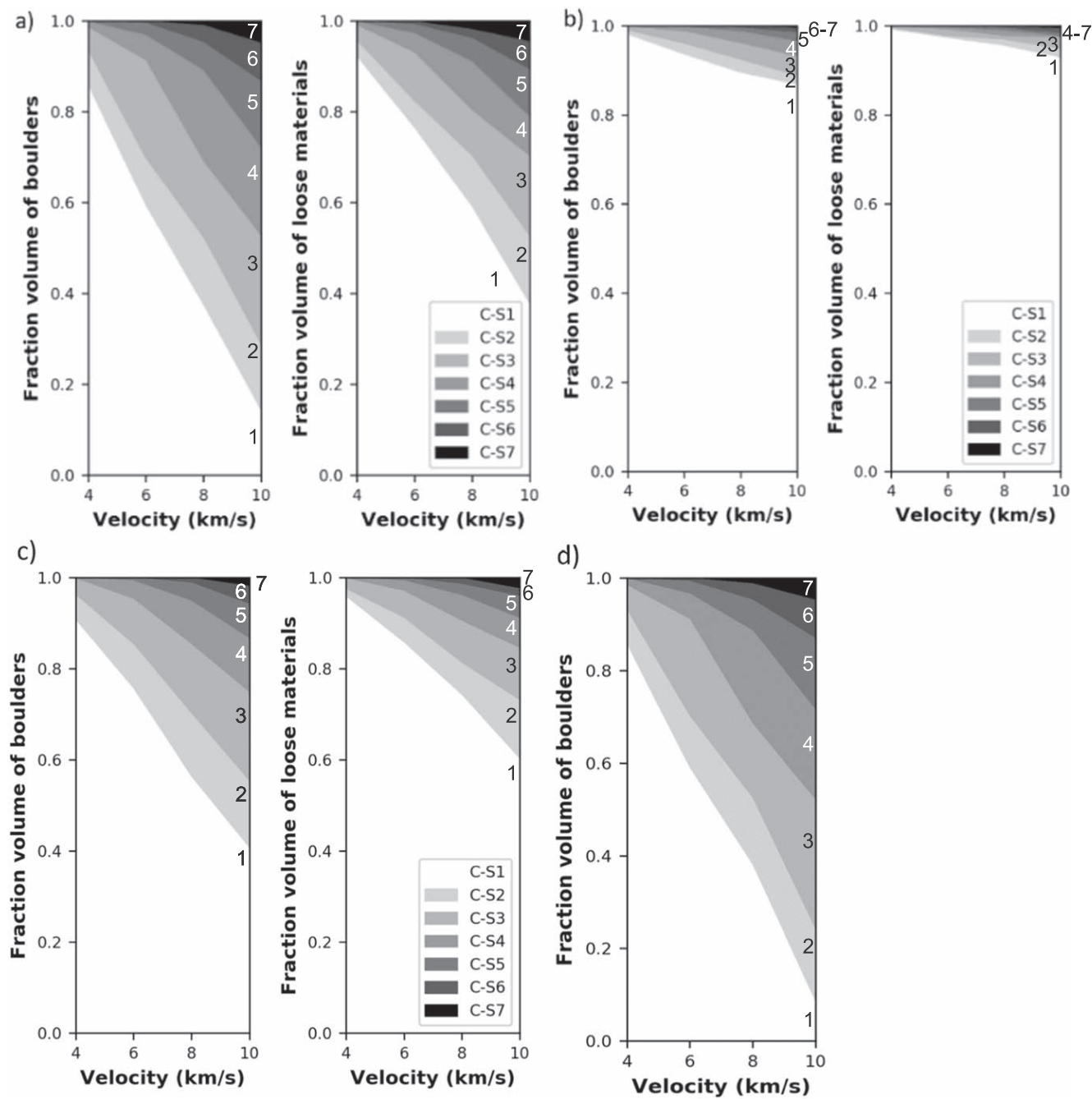

**Figure 5.** Fractions of shocked material. The *y*-axis indicates the apparent fraction of area shocked to a certain stage (Stöffler et al. 2018) as a function of the impact velocity shown on the *x*-axis. The numbers on the right sides of the images indicate shock stages of individual gray polygons, for better identification. In panel (a), a 1600 m impactor collides with a 5 km rubble-pile asteroid with boulders of 10% porosity surrounded by a matrix with 75% porosity. In panel (b), the impactor size is decreased to 800 m in diameter. Panel (c) displays the results of a model where the boulder porosity has been increased to 30%, compared with (a). In panel (d), the matrix consists of void (empty space) and all other parameters are constant, as in (a).

As seen in Figure 7(a), the distribution of pressures in a monolithic asteroid (left panel) of equal porosity is homogeneously distributed and the pressure ranges are similar to the pressures in the rubble-pile asteroid (right panel). Therefore, a monolithic asteroid seems to generate results similar to those of a rubble-pile asteroid if porosity is well-defined within each unit in the asteroid. However, considering a monolithic asteroid, any localized effects and localized pressure amplifications are neglected, and this may not properly represent areas of high shock pressure that may lead to shock-darkening. For the rubble-pile asteroid, we observe that a small apparent fraction of the material experiences shock of shock stage 7, which is not seen in the monolithic asteroid (Figure 7(b)). It is also observed that a larger fraction experiences shock stage 4 in the rubble-pile asteroid compared with the monolithic asteroid.

These small differences are not negligible in evaluating shock-darkening, and we believe that it is essential to consider rubble-pile asteroids in numerical simulations.

Regarding the efficiencies for shock-darkening, the efficiency decreases from 3.3 for the rubble-pile asteroid to 2.84 for the monolithic asteroid considering a large impactor (1600 m), and from 3.22 to 2.7 considering a smaller impactor

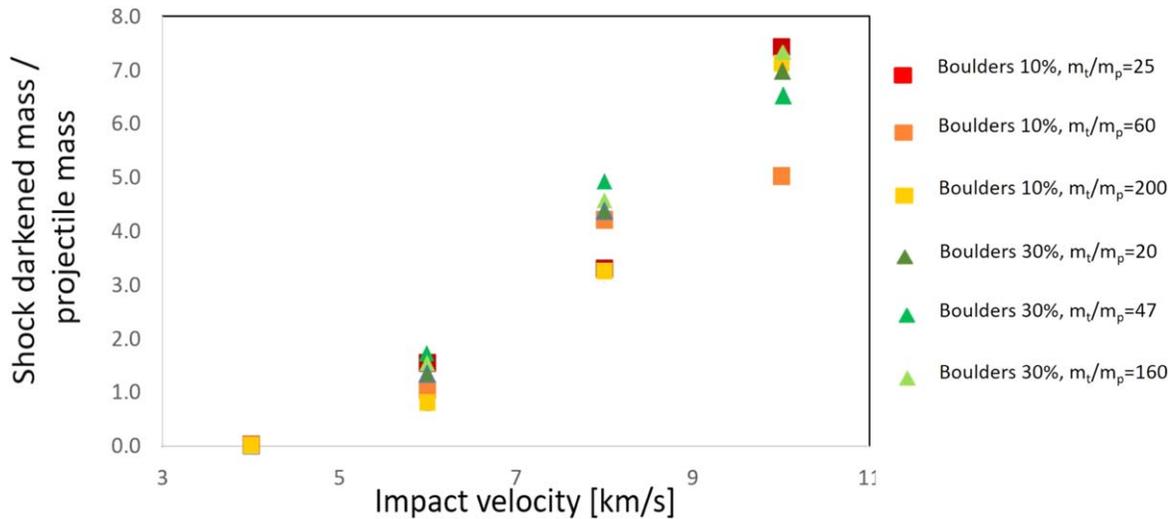

**Figure 6.** Efficiency of shock-darkening as a function of impact velocity. Efficiency is given by the shock-darkened mass normalized to the projectile mass. Square symbols represent simulations with boulder porosities of 10% within the rubble-pile asteroid, and triangular symbols represent boulders with 30% porosity. The ratio of target mass to projectile mass ($m_t/m_p$) corresponds to different impactor sizes. Thus, the respective values for $m_t/m_p$ are: 25 for a 1600 m impactor, 60 for a 1200 m impactor, and 200 for an 800 m impactor.

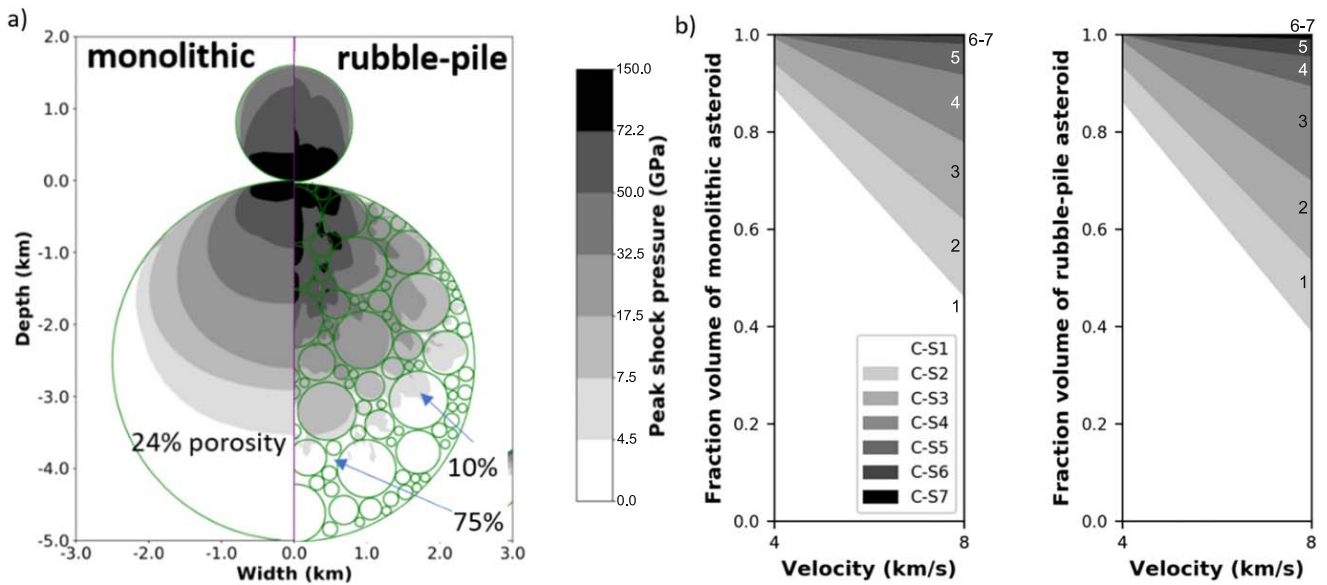

**Figure 7.** (a) Peak shock pressure distribution for a monolithic (left panel) and rubble-pile asteroid (right panel). Tracers are set back to their initial position for better visibility. (b) Apparent fraction of material experiencing certain shock stages (Stöffler et al. 2018) for a monolithic asteroid (left) and a rubble-pile asteroid (right). The numbers in the right part of the images indicate shock stages of individual gray polygons for better identification.

(800 m), both scenarios with an impact velocity of $8\,\mathrm{km\,s^{-1}}$. These differences in efficiency lead to a decrease in shock-darkened masses corresponding to half the mass of one projectile. For an impactor velocity of $4\,\mathrm{km\,s^{-1}}$, the efficiency converges to zero for both rubble-pile and monolithic models. Consequently, the decrease in efficiency considering a monolithic asteroid also supports the necessity to use a rubble-pile asteroid in the model.

### 3.6. Ejecta Analysis

In order to determine how much of the asteroid mass and shock-darkened material is retained or ejected to surrounding space, we investigated the material ejection in detail.

The representative velocity at which material escapes the asteroid was determined to be $0.77\,\mathrm{m\,s^{-1}}$ on average. The details of its calculation are described in the Methods section. Figure 8 presents the results of the analysis of the ejected material given as the ratio of ejected material mass to the mass of the projectile. The ratios of target to projectile mass are given in the legend. A projectile radius of 1600 m leads to a ratio of 25, a radius of 1200 m to 60, and a radius of 800 m to 200 considering a 5 km asteroid target with boulders of 10% porosity surrounded by fine material of 75% porosity. By increasing the boulder porosity to 30%, the mass of the asteroid decreases due to lower bulk densities, and therefore the mass ratios decrease to 20, 47, and 160, respectively. Figure 8(a) shows the total mass of the ejected material normalized to the projectile size as a function of impact velocity, similarly to Figure 6.

We observe that, for the higher impact velocities and bigger projectiles (1600 and 1200 m), almost the entire asteroid is destroyed and all its mass is ejected. With the largest impactor, almost the entire asteroid material is ejected already at 6 km s$^{-1}$ (see red square symbols). At lower impact velocities, only with the largest impactor is a significant amount (about 50%) of material ejected. An increase in boulder porosity has a particularly strong negative effect on the amount of escaping mass, with only a small percentage between 1 and 8% of the total asteroid mass being ejected. The small amount of ejected material is due to the decrease in crater size, although the ejecta would still be fast enough to escape.

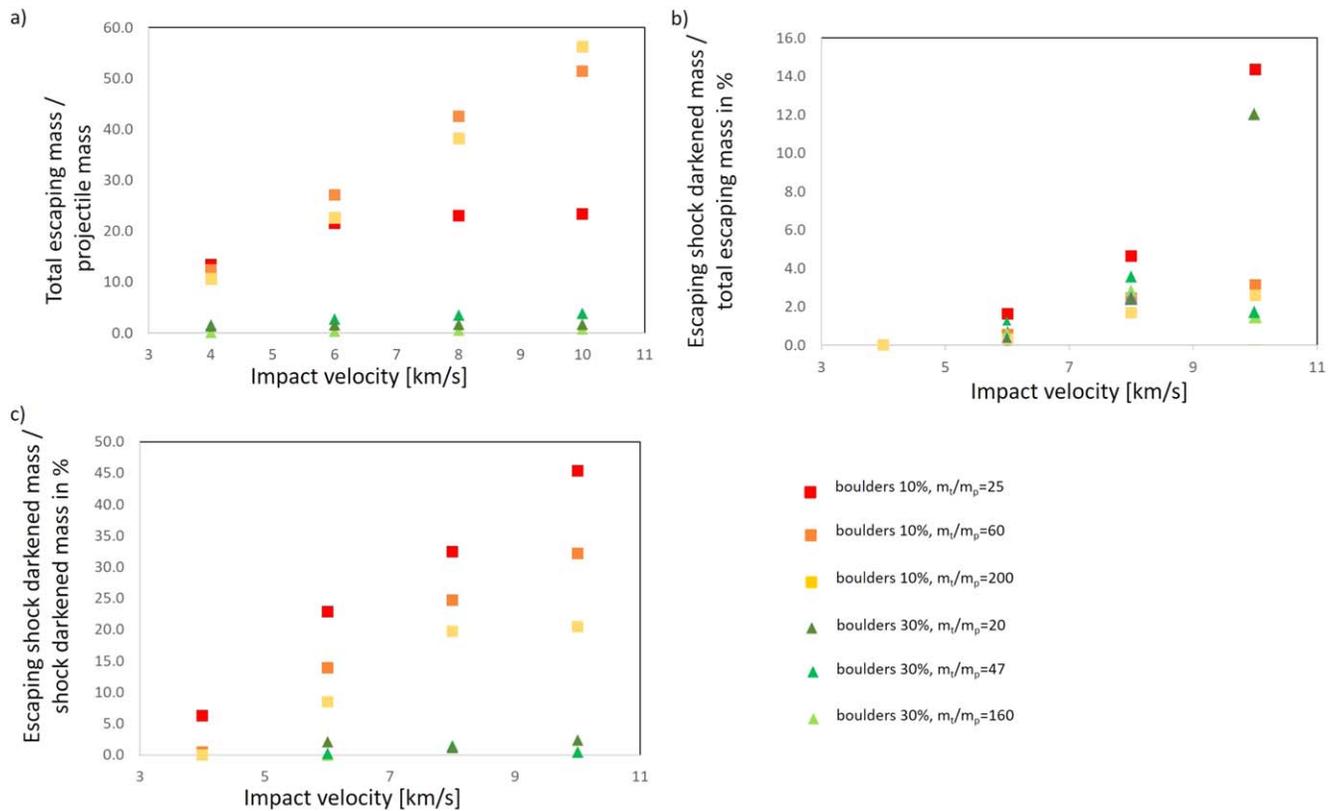

**Figure 8.** Quantification of ejected shock-darkened material as a function of impact velocity for different collision scenarios. Square symbols represent rubble-pile asteroids with boulder porosity of 10% and triangular symbols with boulder porosity of 30%. The fine material has a porosity of 75%. We also indicate different mass ratios of asteroid to impactor through color coding of symbols. (a) Total ejected mass of the asteroid normalized to the projectile mass. (b) Ejected shock-darkened mass as a percentage of the total ejected mass. (c) Ejected shock-darkened mass as a percentage of the total shock-darkened mass of the asteroid.

Generally, the ejection efficiency increases with larger impact velocities and larger impactors. The increase is linear with impact velocity. Of particular interest for this study is how much of the escaping material is shock-darkened. In Figure 8(b), we show the percentage of escaping shock-darkened mass with respect to the total escaping mass as a function of impact velocity for different rubble-pile asteroid settings. We observe that, for almost all collision scenarios, only a small (less than 6%) portion of the total escaping mass is shock-darkened (Figure 8(b)). Only for the largest impact velocity and largest impactor with the entire body being disrupted is a significant increase in ejected shock-darkened material (∼14% of total escaping mass) observed.

In a last step, we look at the percentage of ejected shock-darkened mass with respect to the total shock-darkened mass (Figure 8(c)). It can be observed that, for the most energetic impacts, between 15% and 45% of the shock-darkened mass in the rubble-pile asteroid is ejected. The proportion of ejected shock-darkened mass increases linearly with impact velocity and is largest for the smallest target/projectile mass ratios.

Increasing the boulder porosity has a significant dampening effect on shock-darkened material ejection, with hardly any ejected shock-darkened material.

To summarize, although a significant amount of material is escaping the asteroid, only a small portion of the escaping material is shock-darkened, and the ejection process significantly depends on features of the asteroid structure, such as the boulder porosity, and impact parameters, such as impact velocity and impactor size.

## 4. Discussion

Our results offer insights into the distribution of impact-induced shock pressures within rubble-pile asteroids. The shock wave reflection from a higher-impedance material (Wünnemann et al. 2008; Moreau et al. 2018) is responsible for the pressure increase at the boulder/fine material interface relative to the shock wave direction. Although porosity is known to absorb more energy and to affect the decay of the shock wave (energy to crush pores; Wünnemann et al. 2006), the higher the impedance contrast between a more porous fine material and a less porous boulder, the greater the contrast of pressures at the same impact velocity (Moreau et al. 2018).

We have to consider that lower impact velocities lead to a significant decrease in peak shock pressures, which do not reach the pressure ranges in which shock-darkening may occur (considering pressure ranges between 40 and 60 GPa and above 150 GPa). Smaller impactor sizes (half the diameter), with all other parameters kept constant (boulder and loose material porosity), also lead to a decrease in pressure, although the effect of impact velocity is stronger than that of impactor size.

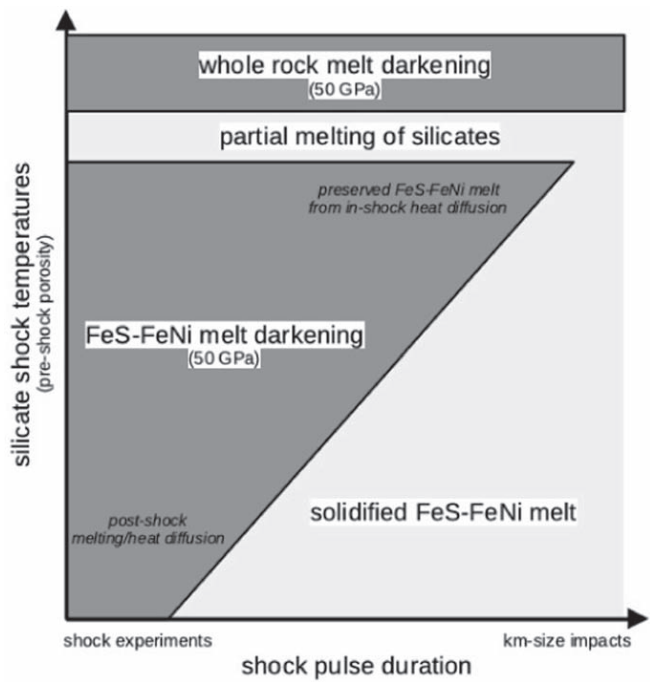

**Figure 9.** Shock-darkening at a given pressure regime happens in specific conditions, which are illustrated in the graphic by dark gray polygons depicting a correlation between the timescale of impact events driving shock pulse duration and the pre-shock porosity of the silicates driving post-shock temperatures. The two dark gray areas are two hypothetical darkening-enabling conditions with metal and iron sulfide melt migration or whole rock melting. A transitional zone of partial melting of silicates inhibits the migration of a metal and iron sulfide melt, and separates the two dark gray polygons. An impact on a large but nonporous body as indicated on the horizontal axis is likely to produce less shock-darkened material, because of FeS and FeNi metal rapidly cooling into surrounding silicates, due to heat conduction, while an impact on a smaller and more porous body may result in more prominent shock-darkening.

The results show that the abundance of high levels of shock metamorphism (>50 GPa or C-S5) in ordinary chondrites requires impacts at high velocities (e.g., 8–10 km s$^{-1}$), or alternatively, large projectiles, to offer a fraction of material exceeding 20% of target volumes shocked to C-S5, C-S6, and C-S7 stages. The current statistics for the ordinary chondrite collection (e.g., Britt & Pieters 1991; Bischoff et al. 2006) indicate that shock-darkened chondrites account for 13%–15% of ordinary chondrite falls, giving evidence for statistically significant presence of shock-darkened material. Rubble-pile asteroids, which show higher porosities (Britt et al. 2002), cannot readily record high shock metamorphism, because highly porous material absorbs a lot of energy due to pore crushing (Güldemeister et al. 2013) and lack highest shock stages (Figure 7). This goes against using monolithic-like numerical models to represent rubble-pile asteroids, although the overall pressure distribution pattern is very similar. Our results also show that the collision events disrupting, or partially fragmenting, parent bodies may originate from impact velocities approaching 10 km s$^{-1}$ (Figures 3 and 8), which is higher than today's average impact velocities in the main asteroid belt (Bottke et al. 1994; Vedder 1998; O'Brien et al. 2011) but not unlikely for the early solar system.

However, it has also been demonstrated that higher porosities require less pressure to possibly display effects equivalent to higher shock stages (Moreau et al. 2019a). An increase in porosity to 15%–30% leads to a decrease in pressures required for shock-darkening by ~10 GPa. We observe that an increase in porosity from 10% to 30% with associated decrease in required pressure leads to an increase in molten material volume by a factor of 1.5. This volume increase is compensated by an overall lower initial material density, and thus, the mass normalized efficiency remains similar to the low-porosity case (see Figure 6). Thus, we can conclude that the effect of pore crushing in the more porous material compensates for the increase in shock-darkening volume with effect on mass-normalized shock-darkening efficiency.

In Moreau & Schwinger (2021), shock-darkening caused by iron sulfide (FeS) or FeNi melt migration is investigated at low initial porosities, and is further discussed in terms of shock pulse duration. They state that, at long shock pulses (corresponding to large-scale collisions), submillimeter mineral grain sizes, and low pre-shock porosities of silicates, the shock-darkening may not happen in 40–60 GPa range, because the heat required to melt the FeS and FeNi-metal at release is rapidly conducted away into surrounding colder silicates during the shock pulse duration. Thus, the FeS and FeNi metal cools to surrounding silicate temperatures before pressure release (Figure 9). However, if the pre-shock porosities increase, the resulting temperature conditions lead to preservation of an FeS-FeNi melt after release of the shock and enable iron sulfide melt migration or whole rock melting, causing shock-darkening as indicated by dark gray polygons in Figure 9. A transitional zone of partial melting of silicates inhibits the migration of a metal and iron sulfide melt, and separates the two dark gray polygons. Thus, the scale of impact and associated shock pulse duration theoretically further constrains the conditions for shock-darkening toward higher-porosity materials. One could further hypothesize that ejected elements are more likely shock-darkened because shock pulses in fragmented and ejected material may be shortened compared with material remaining on the impacted body, where the principal shock wave still propagates.

In this work, we also neglected the contribution of plastic work, which may have a significant effect on the molten and shock-darkened material volume. For example, Manske et al. (2021) and Kurosawa & Genda (2018) show that, for impact velocities lower than 10 km s$^{-1}$, plastic work is not negligible, which would lead to larger melt volumes; hence, our values have to be considered to be lower estimates.

Another relevant aspect in investigating shock-darkened material is the possible ejection of the materials in order to explain the abundance of ordinary chondrites of various shock stages. A collision must somehow be disruptive to match our observation of pressure distributions within ejected material and correlate with recovered ordinary chondrite meteorites in our collections. Therefore, we look at the ejection process and analyze the ejected material and its shock-darkened portion. The collision scenario plays a significant role in how much material is ejected and how much is actually shock-darkened. Here, we can show again that larger impact velocities as well as larger impactor sizes are required. The effect of asteroid porosity plays a significant role in the ejection processes, as well as in the shock-darkening within the asteroid. For example, increasing the boulder porosity from 10% to 30% results in an effective way to reduce the amount of ejected asteroid fragments as a result of asteroid collision events.

Both of these effects—ejection processes as well as shock-darkening efficiency—would further be influenced by varying additional parameters for the impactor, such as impactor porosity or impact angle, which we kept constant. An increase in impactor porosity or impact angle would mostly reduce the peak pressure distribution and hence the shock-darkened material mass and amount of escaping material.

## 5. Conclusions

Shock-darkening happens during collisions with chondritic S-complex rubble-pile asteroids. However, reaching required high levels of shock metamorphism requires impacts with high velocities (8–10 km s$^{-1}$), or alternatively, large projectiles, in order to shock-darken at least 20% of target mass. The shock-darkened asteroid mass fraction ranges from 1.5 to 3 times the impactor mass.

We predict that up to 90% of the asteroid mass can escape, although only less than 6% of the escaping mass is effectively shock-darkened. The escaping mass is strongly affected by the initial porosity of asteroid boulders, which can reduce the percentage of ejected/total shock-darkened mass from 15%–45%, for 10% boulder porosity, to less than 5%, for 30% boulder porosity.

These observations suggest that only a small fraction of asteroids classified as C-complex can be shock-darkened S-complex asteroids of ordinary chondrite composition. Peak shock pressures in monolithic and rubble-pile asteroids are similar; however, it is not possible to detect in monolithic asteroids localized shock effects significantly contributing to shock-darkening. Consequently, impact models with rubble-pile asteroids give a more realistic shock pressure distribution.

We gratefully acknowledge the developers of iSALE-2D, including Gareth Collins, Kai Wünnemann, Dirk Elbeshausen, Boris Ivanov, and Jay Melosh. Some plots in this work were created with the pySALEPlot tool written by Tom Davison. The iSALE-Dellen version was used in this work (Collins et al. 2016). We thank Menghua Zhu and a second, anonymous reviewer for constructive comments on the manuscript.

This work was supported by the Academy of Finland, project Nos. 293975 and 335595, the European Regional Development Fund, the Mobilitas Pluss programme (grant No. MOBJD639), and the NASA Solar System Exploration Research Virtual Institute Center for Lunar and Asteroid Surface Science, and it was conducted within institutional support RVO 67985831 of the Institute of Geology of the Czech Academy of Sciences. R.L. appreciates funding from the European Union's Horizon 2020 research and innovation program, NEO-MAPP, grant agreement No. 870377.

## Supplementary Data File

Full-resolution figures and iSALE model input and output data files can be found at doi:10.5281/zenodo.6470008.

## ORCID iDs

Nicole Güldemeister ● https://orcid.org/0000-0001-8418-6831
Juulia-Gabrielle Moreau ● https://orcid.org/0000-0002-1434-4934
Tomas Kohout ● https://orcid.org/0000-0003-4458-3650
Robert Luther ● https://orcid.org/0000-0002-0745-1467
Kai Wünnemann ● https://orcid.org/0000-0001-5423-1566